\documentclass[11pt,a4paper]{article}
\usepackage{float}
\usepackage{multirow}
\usepackage{graphics}
\usepackage[cp1252,utf8]{inputenc}
\usepackage{hyperref}
\usepackage{graphicx} 
\usepackage{amsmath}
\usepackage{amsfonts}
\usepackage{amssymb}
\usepackage{bigints} 
\usepackage{relsize} 
\usepackage[top=1in, bottom=2cm, left=2cm, right=2cm]{geometry}
\usepackage{authblk}

\usepackage{changepage} 

\usepackage{afterpage}

\usepackage{placeins}

\title{\bf Bulk geometry from entanglement entropy of CFT}
\author[a]{\bf  Ashis Saha \thanks{sahaashis0007@gmail.com, ashisphys18@klyuniv.ac.in}}
\author[b]{\bf Sourav Karar
\thanks{sourav.karar91@gmail.com}}
\author[c]{\bf Sunandan Gangopadhyay \thanks{sunandan.gangopadhyay@gmail.com, sunandan.gangopadhyay@bose.res.in}}

\affil[a]{\textit{Department of Physics, University of Kalyani, Kalyani 741235, India}}
\affil[b]{\textit{Department of Physics, Government General Degree College, Muragachcha, Nadia, India}}
\affil[c]{\textit{Department of Theoretical Sciences,
 S.N.~Bose National Centre for Basic Sciences,}
\textit{JD Block, Sector-III, Salt Lake, Kolkata 700106, India}}

\date{}

\begin{document}

\maketitle

\begin{abstract}
\noindent In this paper, we compute the exact form of the bulk geometry emerging from a $(1+1)$-dimensional conformal field theory using the holographic principle. We first consider the $(2+1)$-dimensional asymptotic $AdS$ metric in Poincare coordinates and compute the area functional corresponding to the static minimal surface $\gamma_A$ and obtain the entanglement entropy making use of the holographic entanglement entropy proposal. We then use the results of the entanglement entropy for $(1+1)$-dimensional conformal field theory
on an infinite line, on an infinite line at a finite temperature and on a circle. Comparing these results with the holographic entanglement entropy, we are able to extract the proper structure of the bulk metric. Finally, we also carry out our analysis in the case of  $\mathcal{N}=4$ super Yang-Mills theory and obtain the exact form of the dual bulk geometry corresponding to this theory. The analysis reveals the behavior of the bulk metric in both the near boundary region and deep inside the bulk. The results also show the influence of the boundary UV cut-off ``$a$" on the bulk metric. It is observed that the reconstructed metrics match exactly with the known results in the literature when one moves deep inside the bulk or towards the turning point.
\end{abstract}

\clearpage

\section*{Introduction}

One of the most interesting and difficult challenge that theoretical physics has come across over the past few decades is to reconcile the general theory of relativity with quantum mechanics. It sounds quite obvious to go for the standard prescription of field quantization to formulate the quantum version of Einstein's general relativity, but unfortunately this programme runs into all sorts of trouble due to the dynamic background involved in Einstein's theory. This has led us to believe that the correct route leading to a complete theory of quantum gravity must be fundamentally different from the route taken for the other known fundamental interactions. This philosophy has led to the holographic idea which says that a gravitational theory has a dual picture in terms of a non-gravitational quantum field theory living on a lower dimensional spacetime, which is the boundary of the spacetime where the gravitational theory lives. Over the past couple of decades great progress has been made in our understanding of quantum theory of gravity, thanks to the remarkable AdS/CFT (gauge/gravity) correspondence \cite{maldacena, maldacena2}, which is intrinsically a non-perturbative approach to finding a quantum gravity theory.

The gauge/gravity correspondence realizes the holographic principle \cite{witten, susskind} in the sense that the information about states in the higher dimensional gravitational system is correlated with the states of the ordinary quantum field theory (without any gravitational degrees of freedom) existing in the lower dimensional manifold. In other words, each observable in the non-gravitational quantum field theory corresponds to some observable of the gravitational theory in the bulk. The duality basically connects a weakly coupled theory with a strongly coupled theory, thus opening a window for exploring a strongly coupled quantum field theory with the help of a weakly coupled gravitational theory or vice versa. The remarkable dual nature of this conjecture has helped us to understand various paradoxes of general relativity, namely, black hole information paradox \cite{tasi}, gravitational singularity \cite{singularity, singularity2}, inflation theory \cite{inflation}, the origin of Hawking radiation \cite{hawkingrad}, to name a few. The correspondence has also led to fundamental results in quantum theory, namely, entanglement in quantum field theory, complexity in quantum field theory \cite{complexity, complexity2}, energy loss of a Brownian particle in quark-gluon plasma \cite{brownian}, to name a few, from the well-established framework of general relativity.

Entanglement entropy (EE) is a well-studied subject in quantum mechanics \cite{nielsen}. The prescription to calculate EE of a quantum field theory is known as `replica trick'. For a 
$(1+1)$-dimensional quantum field theory with conformal group symmetry (CFT), the exact results of EE has been calculated in \cite{Calabrese} for a subsystem defined on various topologies, for instance CFT on a finite strip or CFT on a circle. In higher dimensional ($d>2$) quantum field theory, it becomes notoriously difficult to calculate exact results of EE. Remarkably the holographic principle together with the famous Bekenstein-Hawking ``Area law" \cite{Arealaw},  \cite{BCH} is able to reproduce the EE results available in CFT \cite{gge}, \cite{heepr}. The basic principle of holographic entanglement entropy (HEE) proposed in \cite{RT_prl} states that the EE of a subsystem $(A)$ belonging to a ($d$+1)-dimensional CFT living in the boundary manifold ($\partial \mathcal{M}$) corresponds to the area of the $d$-dimensional minimal surface foliated into a ($d$+2)-dimensional static bulk spacetime ($\mathcal{M}$) where the gravity theory lives. This prescription to calculate the HEE \cite{RT_jhep, RT2} also emphasises the fact that in case of static, asymptotically $AdS$ spacetime equal time foliation produces the surfaces with minimal area. However, in case of a time-dependent and non-static spacetime this prescription does not go through and can be overcomed by introducing the concept of ``minimal surface of maximal slice" to make the prescription covariant \cite{HRT}.  In the context of the HEE proposal, it is worth exploring how the bulk spacetime geometry emerges holographically from a CFT using the exact results for the EE of a subsystem living in the mentioned. In this paper, we investigate the problem of extracting the exact form of the bulk metric through the exact results of EE of a CFT. 
Studies along this direction have been carried out earlier. For instance, 
in \cite{raamsdonk1, raamsdonk2}, Einstein's equation in $AdS$ space were obtained in a perturbative approach. Another nice approach to obtain the bulk metric can be seen in \cite{hamms}-\cite{bilson2}. In \cite{hamms2}, the bulk spacetime metric has been obtained numerically. In \cite{boer}, the length of the bulk curves were obtained using boundary data. It is a well known fact that the EE can be obtained by constructing the reduced density matrix of the concerned subsystem. Hence it is worth asking how the reduced density matrix of the quantum system is holographically connected with the bulk. This was investigated in \cite{raamsdonk3}. In \cite{haehl} the CFT is perturbed with a proper scalar operator which leads to the equivalence of gravity with field propagator in AdS background. Further, the relationship between the correlation functions of a scalar CFT with the correlation functions of self-interacting scalar fields in AdS was established in \cite{koushikray} also the role of renormalization in context of the bulk reconstruction was studied in \cite{solodukhin}. In this paper, we follow the approach in \cite{Bilson} to reconstruct the bulk geometry from the results of the EE of a $(1+1)$-dimensional CFT on an infinite line, on an infinite line at a finite temperature and on a circle. Finally, we reconstruct the bulk geometry for the $\mathcal{N}=4$ super Yang-Mills theory living in $3+1$-dimensions. A crucial input in our analysis is the holographic principle.

The paper is organized as follows. In section \ref{sec1}, we have discussed the basic formalism on which the subsequent analysis rests. In section \ref{linecft}, we obtain the exact bulk metric 
holographically dual to a $1+1$-dimensional CFT on an infinite line. In section \ref{sec3}, we holographically reconstruct the 
dual bulk metric for a $1+1$-dimensional CFT on an infinite line at a finite temperature. In section \ref{sec4}, we extract the exact structure of the bulk geometry for a $1+1$-dimensional CFT on a circle. In section \ref{sec5}, we use the holographic principle to obtain the exact bulk geometry for  $\mathcal{N}=4$ super Yang-Mills theory in $3+1$-dimensions. We conclude in section \ref{conclusion}.

\section{Basic formalism}\label{sec1}
In this section, we briefly discuss the formalism to obtain the exact form of the bulk geometry on the boundary of which lies the CFT. The holographic principle states that the entanglement entropy $S_{A}$ of a static subsystem $A$ in $(d+1)$-dimensional CFT can be obtained from a $d$-dimensional static minimal surface $\gamma_{A}$ in the $(d+2)$-dimensional bulk, the boundary of $\gamma_{A}$ being given by the $(d-1)$-dimensional manifold $\partial\gamma_{A}$. By a static subsystem, we mean a subsystem in the CFT considered at a fixed time. The trick to reconstruct the bulk geometry is to use this principle together with the CFT result for the EE of the static subsystem living in the $(d+1)$-dimensional CFT. The static subsystems $A$ considered in the subsequent analysis are CFT on the infinite line, CFT in a circle and CFT on the infinite line at a finite temperature. The EE of these theories are known from field theoretical considerations. The holographic entanglement entropy (HEE) is given by the Bekenstein-Hawking formula 
\begin{eqnarray}\label{ent_bh}
	S_{A} = \frac{Area(\gamma_{A})}{4G_{N}^{(d+2)}}
\end{eqnarray}	
where $Area(\gamma_{A})$ is the area of the static minimal surface $\gamma_{A}$ and $G_{N}^{(d+2)}$ is the Newton's gravitational 
universal constant in $(d+2)$-dimensions.

\noindent We start our analysis by considering the asymptotically $AdS$ planar metric in Poincare coordinates in $(2+1)$-dimensions \cite{Braga}
\begin{eqnarray}\label{ads_m}
    ds^2= \frac{R^2}{z^2} \Bigg( -h(z)dt^2+f(z)^2 dz^2 + dx^2 \Bigg)
    \end{eqnarray}
where $z \geq 0$ denotes the bulk coordinate and $z=0$ defines the boundary of the bulk where the CFT lives. The choice of a static spacetime ansatz in eq.(\ref{ads_m}) can be justified from the EE results of CFT that we have considered in our analysis. A stationary, non-static spacetime (for example rotating BTZ) is dual to a CFT on a circle at a finite temperature $\beta^{-1}$ with a potential $\Omega$ associated with the momentum of the CFT, where the CFT is described by the density matrix $\rho = \exp ({-\beta H+\beta \Omega P})$ \cite{HRT}. Equivalently, this would correspond to the inverse temperatures of the left and right moving modes 
$\beta_{\pm} = \beta (1 \pm \Omega)$ appearing in the expression for the EE of the CFT. 

\noindent Now the area functional of the above geometry reads (setting $dt=0$) 
\begin{eqnarray}\label{area}
 Area(\gamma_A)= R \int dx \frac{\sqrt{[z^\prime f(z)]^2 +1}}{z}~, ~ z^\prime =\frac{dz}{dx}~.
\end{eqnarray}
To obtain the static minimal surface $\gamma_A$, we need to minimize the area functional above. To do this, the first step is to compute the Hamiltonian which reads
\begin{eqnarray}
\mathcal{H}=z^\prime \frac{d\mathcal{L}}{dz^\prime}-\mathcal{L}=-\frac{1}{z\sqrt{[z^\prime f(z)]^2 +1}}
\end{eqnarray}
where the Lagrangian ($\mathcal{L}$) is given by $\mathcal{L} (z, z^\prime,x) =  \frac{\sqrt{[z^\prime f(z)]^2 +1}}{z}~.$ 
Since the Hamiltonian  has no explicit $x$ dependence, hence it is a constant. To determine this constant, we use the fact that $\frac{dz}{dx}|_{z=z*}=0$ where $z=z_*$ denotes the turning point of the minimal surface  $\gamma_A$ in the bulk. This yields
\begin{eqnarray}\label{derr_bulk}
\frac{dz}{dx}= - \frac{\sqrt{z_* ^2-z^2}}{z f(z)} ~.
\end{eqnarray}

Substituting eq.(\ref{derr_bulk}) in eq.(\ref{area}), we get
\begin{eqnarray}\label{area*}
Area[\gamma_A(z_*)]= R \int dz \frac{z_* f(z)}{z\sqrt{z_* ^2-z^2}}~.
\end{eqnarray}    
Now substituting eq.(\ref{area*}) in eq.(\ref{ent_bh}, we obtain the HEE to be 
\begin{eqnarray}\label{S_ee}
S_{A} =\frac{Area[\gamma_{A}(z_*)]}{4G_{N}^{(3)}}       =\frac{1}{4G_{N}^{(3)}} \int dz \frac{z_* f(z)}{z\sqrt{z_* ^2-z^2}}~.
\end{eqnarray}
Our aim in this paper is to obtain $f(z)$ by comparing this result with the CFT result for the EE employing the approach in \cite{Bilson}. The metric coefficient of $dt^2$ can then be obtained by substituting $f(z)$ in Einstein's field equations of general relativity with a negative cosmological constant
\begin{eqnarray}\label{adsE}
G_{\mu \nu}= \mathcal{R}_{\mu \nu}-\frac{1}{2} g_{\mu \nu} \mathcal{R} +\Lambda g_{\mu \nu} = 8\pi G T_{\mu \nu}~, \qquad \Lambda = -\frac{1}{R^2}~.
\end{eqnarray} 


\section{(1+1)-dimensional CFT on an infinite line}\label{linecft} 
In this section we would like to obtain the bulk metric corresponding to the result for the EE of the $(1+1)$-dimensional CFT on the infinite line. This reads \cite{Calabrese}
\begin{eqnarray*}
S_{EE}(l)= \frac{c}{3}\log\bigg(\frac{l}{a} \bigg)
\end{eqnarray*}
where $c=\frac{3R}{2G_N}$ is the central charge representing the degrees of freedom in the CFT \cite{brownhenn}, $l$ denotes the length of the subsystem and $a$ is the lattice spacing or the UV cut-off introduced in the boundary field theory so that the information density is bounded \cite{witten}. One can rewrite the above expression using the definition of central charge $c$ as
\begin{eqnarray}\label{cft1}
S_{EE}(l)= \frac{2R}{4G_N} \log\bigg(\frac{l}{a}\bigg)~.
\end{eqnarray}
To obtain the bulk geometry corresponding to this CFT, we write down the area functional (\ref{area}) once again
\begin{eqnarray}\label{area_l}
 Area[\gamma_A] &=& R \int_{-l/2}^{+l/2} dx \frac{\sqrt{[z^\prime f(z)]^2 +1}}{z} \nonumber \\
 &=& 2R \int_{0}^{l/2} dx \frac{\sqrt{[z^\prime f(z)]^2 +1}}{z} \nonumber \\
 &=& 2R \int_{a}^{z_*} dz \frac{z_* f(z)}{z\sqrt{z_* ^2-z^2}} \nonumber \\
 &=& Area[\gamma_A(z_*)] \equiv \mathcal{A}_{l}(z_*)
\end{eqnarray}   
where in the third line we have used eq.(\ref{derr_bulk}).\\ 
In the third line of the above expression we have introduced a cut-off surface at $z=a$ (which in turn makes the bulk coordinate $z>a$) to regularize the area functional. This cut-off is directly related with  the lattice spacing of the boundary field theory \cite{witten}.\\
\noindent The HEE of the subsystem $A$ of length $l$ therefore reads 
\begin{eqnarray}\label{bulk1}
S_{A}= \frac{\mathcal{A}_l(z_*)}{4G_N}~.
\end{eqnarray}
\noindent From eq.(\ref{derr_bulk}), we can also write the length of the subsystem $l$ in terms of the bulk coordinate $z$ as
\begin{eqnarray}\label{sys}
l = 2 \int_{0}^{z_*}\frac{zf(z)}{\sqrt{z_*^2-z^2}} dz ~.
\end{eqnarray}
It is to be noted that we have not put the UV cut-off ``$a$" in the lower limit of the above integral since the integral has no divergence. 

\noindent To proceed further, we would like to have a relation between $z_*$ and $l$. To get this relation, we note that according to the holographic principle
\begin{eqnarray}\label{equate}
\frac{dS_{EE}(l)}{dl} =  \frac{dS_{A}}{dl}~.
\end{eqnarray}    
\noindent Now using eq.(\ref{cft1}), we have
\begin{eqnarray}\label{ee1}
 \frac{dS_{EE}(l)}{dl}= \frac{2R}{4G_N} \frac{1}{l} 
\end{eqnarray}
and using eq.(s)(\ref{area_l}), (\ref{sys}), we have
\begin{eqnarray}\label{hee1}
\frac{dS_{A}}{dl}=\frac{1}{4G_N} \frac{d\mathcal{A}_l(z_*)}{dz_*} \frac{dz_*}{dl}=\frac{2R}{4G_N} \frac{1}{2z_*}~.
\end{eqnarray}
\noindent Substituting eq.(s)(\ref{ee1}) and (\ref{hee1}) in eq.(\ref{equate}), we get
\begin{eqnarray}\label{lz}
l=2z_*~.
\end{eqnarray}
Substituting eq.(\ref{lz}) in eq.(\ref{cft1}), we can rewrite the expression for EE of the CFT in terms of the bulk coordinate as
\begin{eqnarray}\label{cftz1}
S_{EE}(z_*)= \frac{2R}{4G_N} \log \bigg(\frac{2z_*}{a}\bigg)~.
\end{eqnarray}
On the basis of the holographic principle mentioned earlier, we now equate eq.(s)(\ref{bulk1}) and (\ref{cftz1}) to obtain the expression of the area functional $\mathcal{A}_l(z_*)$ to be
\begin{eqnarray}
\mathcal{A}_l(z_*) = 4G_N S_{EE}(z_*)
\end{eqnarray}      
which in turn yields (using eq.(\ref{area_l}))
\begin{eqnarray}
\frac{4G_N}{2Rz_*} S_{EE}(z_*) = \int_{a}^{z_*} dz \frac{f(z)}{z\sqrt{z_* ^2-z^2}} ~.
\end{eqnarray} 
Setting $\frac{4G_N}{2Rz_*} S_{EE}(z_*)= \mathcal{B}(z_*)$, $f(z)/z = m(z)$, the above equation can be recast as
\begin{eqnarray}
\mathcal{B}(z_*)= \int_{a}^{z_*} dz \frac{m(z)}{\sqrt{z_*^2- z^2}}~.
\end{eqnarray}
The solution to this Volterra first kind (Abel type) integral equation reads \cite{abels, polyanin}
\begin{eqnarray}\label{int.eq}
{m}(z)= \frac{1}{\pi} \frac{d}{dz} \int_{a}^{z} dz_* \frac{\mathcal{B}(z_*)2z_*}{\sqrt{z^2-z_*^2}}~.
\end{eqnarray}
Substituting $m(z)$ and $\mathcal{B}(z_*)$ in the above equation, we get
\begin{eqnarray}\label{result1}
f(z)= \frac{2 z}{\pi}  \frac{d}{dz} \int_{a}^{z} dz_* \frac{\log(2z_*/a)}{\sqrt{z^2-z_*^2}} \equiv \frac{2 z}{\pi} \frac{dI}{dz}~.
\end{eqnarray}
We now proceed to calculate $\frac{dI}{dz}$ in the above expression. This gives
\begin{eqnarray}\label{0001}
\frac{dI}{dz}&=& \frac{\pi}{2z} + \frac{1}{z} \frac{(a/z)}{\sqrt{1-(a/z)^2}} \log 2 - \frac{1}{z}\sin^{-1}(a/z)~.
\end{eqnarray}
Substituting the above result in eq.(\ref{result1}), we get
\begin{eqnarray}\label{f1}
f(z) &=& 1+ \frac{2}{\pi} \frac{(a/z)}{\sqrt{1-(a/z)^2}} \log 2 - \frac{2}{\pi} \sin^{-1}(a/z)~.
\end{eqnarray}
It is interesting to observe that the value of the metric coefficient $f(z)$ is exact upto all orders in $(a/z)$, $(z>a)$. The above expression of $f(z)$ captures the effect of the bulk-boundary UV cut-off `$a$'. The effect arises due to the presence of the lattice spacing in the boundary field theory.\\
In the limit $(a/z)\rightarrow 0$, we get
$$f(z)=1$$
thereby fixing the coefficient of $dz^2$ in the metric (\ref{ads_m}). Hence we have
\begin{eqnarray}\label{line1}
ds^2= \frac{R^2}{z^2} \Bigg( -h(z)dt^2+ dz^2 + dx^2 \Bigg) \equiv  g_{\mu \nu}dx^{\mu} dx^{\nu}~;~ \mu,\nu = 1,2,3~.
\end{eqnarray}
\noindent The metric ($\ref{line1}$)  gives the following Einstein field equations
\begin{eqnarray}
G_{22}&=&-\frac{h^\prime (z)}{2zh(z)}=0\label{ge1}\\
G_{33}&=&-\frac{z h^\prime(z)^2+2h(z)h^\prime (z)-2zh(z)h^{\prime\prime}(z)}{4zh(z)^2}\label{ge2}=0~.
\end{eqnarray}
Solving eq.(\ref{ge1}) gives $h(z)= constant = \mathcal{K}$ which also satisfies eq.(\ref{ge2}).

\noindent The exact form of the bulk geometry in the limit $(a/z)\rightarrow 0$ corresponding to the CFT on an infinite line therefore has the form
\begin{eqnarray}\label{final1}
ds^2= \frac{R^2}{z^2} \Bigg( - \mathcal{K}dt^2+ dz^2 + dx^2 \Bigg)~.
\end{eqnarray}
\noindent This is the well known pure $AdS_3$ metric in Poincare coordinates.



\section{(1+1)-dimensional CFT on an infinite line at a finite temperature}\label{sec3}
\noindent In this section we analyse the subsystem $A$ of length $l$ discussed in the previous section at a finite temperature $T$. The entanglement entropy of the CFT at a finite temperature from field theoretic considerations reads \cite{Calabrese}
\begin{eqnarray}\label{cft2}
S_{EE}(l)= \frac{2R}{4G_{N}} \log \bigg[ \frac{\beta}{\pi a} \sinh\bigg(\frac{\pi l}{\beta}\bigg)\bigg]
\end{eqnarray}
where $\beta=1/T$ represents the temperature of the CFT.

\noindent Differentiating $S_{EE}(l)$ with respect to $l$ gives
\begin{eqnarray}\label{ee2}
 \frac{dS_{EE}(l)}{dl}= \frac{2R}{4G_N} \bigg(\frac{\pi}{\beta}\bigg) \coth\bigg(\frac{\pi l}{\beta}\bigg)~.
\end{eqnarray}
\noindent Substituting eq.(s)(\ref{hee1}, \ref{ee2}) in eq.(\ref{equate}),
we get the length $l$ of the subsystem $A$ in terms of the turning point $z_*$ to be
\begin{eqnarray}\label{sys2}
l = \frac{\beta}{\pi} \coth^{-1}\bigg( \frac{\beta}{2\pi z_*}\bigg)~.
\end{eqnarray}
The expression for the EE (\ref{cft2}) 
can now be recast using eq.(\ref{sys2}) as
\begin{eqnarray}\label{cft*}
S_{EE}(z_*)= \frac{2R}{4G_{N}} \log \bigg[ \frac{\beta}{\pi a} \sinh \bigg(\coth^{-1}\bigg(\frac{\beta}{2\pi z_*}\bigg)\bigg)\bigg]~.
\end{eqnarray}
Substituting the above result in eq.(\ref{S_ee}), we get
\begin{eqnarray}\label{temp}
f(z)= \frac{2}{\pi} z \frac{d}{dz} \bigints_{a}^{z} dz_* \frac{\log \bigg[ \frac{\beta}{\pi a} \sinh \big(\coth^{-1}(\frac{\beta}{2\pi z_*})\big)\bigg]}{\sqrt{z^2-z_*^2}} \equiv \frac{2 z}{\pi} \frac{dJ}{dz}~.
\end{eqnarray}\\
\noindent Now $\frac{dJ}{dz}$ in the above expression reads
\begin{eqnarray}\label{001}
\frac{dJ}{dz} &=& \bigg(\frac{1}{z}\bigg) \frac{(a/z)}{\sqrt{1-(a/z)^2}} \log\bigg[ \frac{2}{\sqrt{1-(2\pi a/\beta)^2}}\bigg] - \bigg(\frac{1}{z}\bigg) \sin^{-1} (a/z) + \bigg(\frac{\pi}{2z}\bigg) \frac{1}{\sqrt{1-(2\pi z/\beta)^2}}\nonumber\\
&& + \frac{1}{z} \tan^{-1} \bigg(\frac{(a/z)}{\sqrt{1-(a/z)^2}}\bigg) - \bigg(\frac{1}{z}\bigg) \frac{1}{\sqrt{1-(2\pi z/\beta)^2}} \tan^{-1}\bigg( \frac{a}{z} \sqrt{\frac{1-(2\pi z/\beta)^2}{1-(a/z)^2}} \bigg)~.
\end{eqnarray}
Using this result in eq.(\ref{temp}), we get
\begin{eqnarray}\label{002}
f(z)&=& \bigg(\frac{2}{\pi}\bigg) \frac{(a/z)}{\sqrt{1-(a/z)^2}} \log\bigg[ \frac{2}{\sqrt{1-(2\pi a/\beta)^2}}\bigg] - \bigg(\frac{2}{\pi}\bigg) \sin^{-1} (a/z) + \frac{1}{\sqrt{1-(2\pi z/\beta)^2}}\nonumber\\
&& + \bigg(\frac{2}{\pi}\bigg) \tan^{-1} \bigg(\frac{(a/z)}{\sqrt{1-(a/z)^2}}\bigg) - \bigg(\frac{2}{\pi}\bigg) \frac{1}{\sqrt{1-(2\pi z/\beta)^2}} \tan^{-1}\bigg( \frac{a}{z} \sqrt{\frac{1-(2\pi z/\beta)^2}{1-(a/z)^2}} \bigg)~.
\end{eqnarray}\\
It is reassuring to note that the above expression for $f(z)$ reduces to eq.(\ref{f1}) in the limit $\beta \rightarrow \infty$ (that is in the zero temperature limit). It is also remarkable to observe that the above expression for the metric coefficient $f(z)$ is exact upto all orders in $a/z$, $(z>a)$.\\
\noindent In the limit $\frac{a}{z}\rightarrow0$, 
eq.(\ref{002}) reduces to 
\begin{eqnarray}
 f(z)=\frac{1}{\sqrt{1-\frac{4\pi^2 z^2}{\beta^2}}}~.
\end{eqnarray}
Now substituting $f(z)$ in eq.(\ref{ads_m}), we get
\begin{eqnarray}\label{temp1}
ds^2= \frac{R^2}{z^2} \Bigg( -h(z)dt^2+ \frac{dz^2}{1-\frac{4\pi^2 z^2}{\beta^2}} + dx^2 \Bigg) \equiv  g_{\mu \nu}dx^{\mu} dx^{\nu}~;~\mu, \nu =1,2,3.
\end{eqnarray}  
The above metric leads to the following Einstein's field equations
\begin{eqnarray}\label{eg3}
 G_{22}=-\frac{8 \pi^2 z h(z)+\beta^2 h^\prime (z)-4 \pi^2 z^2 h^\prime (z)}{2b^2zh(z)-8\pi^2z^3h(z)}=0
 \end{eqnarray}
\begin{eqnarray}\label{eg4}
G_{33}=\frac{z(4\pi^2z^2-\beta^2)h^{\prime2} (z)-2h(z)[\beta^2 h^\prime (z)+z(4\pi^2 z^2-\beta^2)h^{\prime \prime}(z)]}{4 \beta^2 z h(z)^2}=0.
\end{eqnarray}
Solving eq.(\ref{eg3}), we get  
\begin{eqnarray}\label{h}
h(z)=(Constant)\beta^2 \bigg(1-\frac{4\pi^2 z^2}{\beta^2}\bigg)=\mathcal{C}\bigg(1-\frac{4\pi^2 z^2}{\beta^2}\bigg)
\end{eqnarray}
which satisfies eq.(\ref{eg4}). Substituting $h(z)$ in the metric (\ref{temp1}), we obtain the final form of the bulk geometry corresponding to the CFT on an infinite line at a finite temperature $T$ to be
\begin{eqnarray*}
ds^2= \frac{R^2}{z^2} \Bigg[ -\mathcal{C}\bigg(1-\frac{4\pi^2 z^2}{\beta^2}\bigg)dt^2+ \frac{dz^2}{1-\frac{4\pi^2 z^2}{\beta^2}} + dx^2 \Bigg]~.
\end{eqnarray*}
Using the coordinate transformation $z=\frac{R}{r}$, we can recast the above metric in a familiar form as
\begin{eqnarray}\label{BtZ}
ds^2= -\mathcal{C} (r^2-r_{+}^2)dt^2+ \frac{R^2 dr^2}{(r^2-r_{+}^2)} + r^2 dx^2 
\end{eqnarray}
which is the static BTZ black hole metric \cite{BTZ} with the event horizon at $r_{+}=\frac{2\pi R}{\beta}~.$

\noindent The above analysis represents the fact that the $2+1$-dimensional static BTZ geometry emerges holographically from the $1+1$-dimensional CFT on an infinite line at a finite temperature $T$.

 

\section{(1+1)-dimensional CFT on a circle}\label{sec4} 

In this section we shall obtain the exact form of the bulk geometry corresponding to the result for the EE of the $(1+1)$-dimensional CFT on a circle. From conformal field theoretic considerations, the result for the EE of a subsystem $A$ of length $l$ on a circle reads \cite{Calabrese}
\begin{eqnarray}\label{circle_ee}
 S_{EE}(l)=\frac{2R}{4G_N}\log \bigg[\frac{L}{\pi a} \sin\bigg(\frac{\pi l}{L}\bigg)\bigg]
\end{eqnarray}
where $L$ represents the total length of the circle.
 
\noindent The present situation is quite different from the previous cases due to the presence of circular symmetry. This fact requires to introduce the circular symmetry in the metric ansatz (\ref{ads_m}). To do this, we shall use the basic property of a circle which is
\begin{eqnarray}\label{ang_c}
\bigg(\frac{2\pi}{L} dx \bigg) = d\theta~;~\theta \approx \theta +2\pi ~.
\end{eqnarray}
With the above relation in mind, we take the metric ansatz in the following form
\begin{eqnarray}\label{ads_m_circle}
ds^2= \frac{R^2}{z^2} \Bigg[ -h(z)dt^2+f(z)^2 dz^2 + \bigg(\frac{L}{2\pi}\bigg)^2 d\theta^2 \Bigg]~.
\end{eqnarray}
This leads to the following area functional 
\begin{eqnarray*}
Area[\gamma_A]&=& R \int_{-{\pi l}/L}^{{+\pi l}/L} d\theta \frac{\sqrt{[z^\prime f(z)]^2 +(\frac{L}{2\pi})^2}}{z}\nonumber\\
&=& 2R \int_{0}^{{\pi l}/L} d\theta \frac{\sqrt{[z^\prime f(z)]^2 +(\frac{L}{2\pi})^2}}{z}~;~z^\prime = \frac{dz}{d\theta}~.
\end{eqnarray*}
As before, we once again write down the area functional and subsystem size $l$ in terms of the bulk coordinate $z$. Hence we have
\begin{eqnarray}\label{circle_area_z1}
Area[\gamma_A(z_*)] \equiv \mathcal{A}_{l}(z_*) = 2R \int_{a}^{z_*} dz \frac{z_* f(z)}{z\sqrt{z_* ^2-z^2}}
\end{eqnarray}
\begin{eqnarray}
\label{lll}
l = 2 \int_{a}^{z_*}\frac{zf(z)}{\sqrt{z_*^2-z^2}} dz ~.
\end{eqnarray}
The HEE is therefore given by
\begin{eqnarray}\label{circle**}
S_{A} = \frac{\mathcal{A}_{l}(z_*)}{4 G_N} = \frac{2R}{4G_N} \int_{a}^{z_*} dz \frac{z_* f(z)}{z\sqrt{z_* ^2-z^2}}~.
\end{eqnarray}
Now it is observed from eq.(\ref{circle_ee}) that
\begin{eqnarray}
\frac{dS_{EE}(l)}{dl}=\bigg(\frac{2R}{4G_N}\bigg)\bigg(\frac{\pi}{L}\bigg) \cot\bigg(\frac{\pi}{L}\bigg)
\end{eqnarray} 
which in turn gives using eq.(s)(\ref{equate}), (\ref{lll}) and (\ref{circle**})
\begin{eqnarray*}
l= \frac{L}{\pi} \cot^{-1}\bigg(\frac{L}{2\pi z_*}\bigg)~.
\end{eqnarray*} 
With the above relation in place, we can recast the expression for the EE (\ref{circle_ee}) in terms of the bulk coordinate as 
\begin{eqnarray}
 S_{EE}(z_*)= \frac{2R}{4G_N}\log\bigg[\frac{L}{\pi a} \sin\bigg(\cot^{-1}\bigg(\frac{L}{2\pi z_*}\bigg)\bigg)\bigg]~.
\end{eqnarray}
Substituting the above result in eq.(\ref{area*}), we obtain
\begin{eqnarray}\label{int.eq3}
f(z)= \frac{2}{\pi} z \frac{d}{dz} \bigints_{a}^{z} dz_* \frac{\log \bigg[ \frac{L}{\pi a} \sin \big(\cot^{-1}(\frac{L}{2\pi z_*})\big)\bigg]}{\sqrt{z^2-z_*^2}} \equiv \frac{2z}{\pi} \frac{dK}{dz}~.
\end{eqnarray}
We now proceed to calculate $\frac{dK}{dz}$ in the above expression. This yields
\begin{eqnarray}\label{003}
\frac{dK}{dz} &=& \bigg(\frac{1}{z}\bigg) \frac{(a/z)}{\sqrt{1-(a/z)^2}} \log\bigg[ \frac{2}{\sqrt{1+(2\pi a/L)^2}}\bigg] - \bigg(\frac{1}{z}\bigg) \sin^{-1} (a/z) + \bigg(\frac{\pi}{2z}\bigg) \frac{1}{\sqrt{1+(2\pi z/L)^2}}\nonumber\\
&& + \frac{1}{z} \tan^{-1} \bigg(\frac{(a/z)}{\sqrt{1-(a/z)^2}}\bigg) - \bigg(\frac{1}{z}\bigg) \frac{1}{\sqrt{1+(2\pi z/L)^2}} \tan^{-1}\bigg( \frac{a}{z} \sqrt{\frac{1+(2\pi z/L)^2}{1-(a/z)^2}} \bigg)~.
\end{eqnarray}\\
\noindent Substituting the above result in eq.(\ref{int.eq3}), we get
\begin{eqnarray}\label{004}
f(z)&=& \bigg(\frac{2}{\pi}\bigg) \frac{(a/z)}{\sqrt{1-(a/z)^2}} \log\bigg[ \frac{2}{\sqrt{1+(2\pi a/L)^2}}\bigg] - \bigg(\frac{2}{\pi}\bigg) \sin^{-1} (a/z)+  \frac{1}{\sqrt{1+(2\pi z/L)^2}}\nonumber\\
&& + \frac{2}{\pi} \tan^{-1} \bigg(\frac{(a/z)}{\sqrt{1-(a/z)^2}}\bigg) - \bigg(\frac{2}{\pi}\bigg) \frac{1}{\sqrt{1+(2\pi z/L)^2}} \tan^{-1}\bigg( \frac{a}{z} \sqrt{\frac{1+(2\pi z/L)^2}{1-(a/z)^2}} \bigg)~.
\end{eqnarray}
It is reassuring to note that the above expression for $f(z)$ reduces to eq.(\ref{f1}) in the limit $L \rightarrow \infty$. It is also remarkable to observe that the above expression for the metric coefficient $f(z)$ is exact upto all orders in $a/z$, $(z>a)$.\\
In the limit $\frac{a}{z} \rightarrow 0$, the above expression for $f(z)$ reduces to
\begin{eqnarray}\label{fz}
f(z)=\frac{1}{\sqrt{1+\frac{4\pi^2z^2}{L^2}}}~.
\end{eqnarray}
Now substituting $f(z)$ in eq.(\ref{ads_m_circle}), we get
\begin{eqnarray}
ds^2= \frac{R^2}{z^2} \Bigg[ -h(z)dt^2+ \frac{dz^2}{1+\frac{4\pi^2z^2}{L^2}} dz^2 + \bigg(\frac{L}{2\pi}\bigg)^2 d\theta^2 \Bigg] \equiv g_{\mu \nu} dx^{\mu} dx^{\nu}~;~\mu, \nu = 1,2,3~.
\end{eqnarray}
The above metric leads to the following Einstein's field equations
\begin{eqnarray}\label{g22}
G_{22} = \frac{8\pi^2 z h(z) - L^2 h^{\prime}-4 \pi^2 z^2 h^{\prime}}{2L^2z h(z)+ 8\pi^2 z^3 h(z)} = 0
\end{eqnarray}
\begin{eqnarray}\label{g33}
G_{33} = \frac{-z(L^2+4\pi^2 z^2)h^{2\prime}+h(z) [-2L^2 h^{\prime}(z)+2z (L^2+4\pi^2 z^2)h^{\prime \prime}(z)]}{4L^2 z h^2(z)} = 0~.
\end{eqnarray}
Solving eq.(\ref{g22}), we obtain
\begin{eqnarray*}
h(z)= (Constant) L^2(1+\frac{4\pi^2z^2}{L^2}) = \mathcal{M}(1+\frac{4\pi^2z^2}{L^2})
\end{eqnarray*}
which satisfies eq.(\ref{g33}). The exact form of the bulk geometry corresponding to the CFT on a circle therefore reads
\begin{eqnarray*}
ds^2= \frac{R^2}{z^2} \Bigg[ -\mathcal{M}\bigg(1+\frac{4\pi^2z^2}{L^2}\bigg)dt^2+\frac{dz^2}{1+\frac{4\pi^2z^2}{L^2}} + (\frac{L}{2\pi})^2 d\theta^2 \Bigg]~.
\end{eqnarray*}\\

\noindent We shall rewrite the above metric in a more familiar form by casting it in global coordinates $\frac{2\pi z}{L}\sinh\rho=1$. This leads to
\begin{eqnarray}\label{global}
ds^2=R^2\Bigg(-\mathcal{M}\cosh^2\rho dt^2+d\rho^2+\sinh^2 \rho d\theta^2\Bigg).
\end{eqnarray}
This is the well known pure $AdS_33$ metric in global coordinates.

\section{$\mathcal{N}=4$ super Yang-Mills theory}\label{sec5} 
In this section we shall obtain the exact form of the bulk geometry which corresponds to the result of the EE for $\mathcal{N}=4$ super Yang-Mills theory. The EE of this theory for a strip like subsystem ($l$) reads \cite{RT_prl}
\begin{eqnarray}\label{yangmills}
S_{EE}(l) = \frac{N^2 L^2}{2 \pi a^2} - 2\sqrt{\pi}\bigg[\frac{\Gamma(\frac{2}{3})}{\Gamma(\frac{1}{6})}\bigg]^3 \frac{N^2 L^2}{l^2}
\end{eqnarray}
where $N^2 = \frac{\pi R^3}{2 G_{N}}$ and $L$ represents the width of the subsystem strip.

\noindent The present scenario is quite different from the previous cases as it is a higher dimensional theory. In this case, we take the metric ansatz as

\begin{eqnarray}\label{ymmetric}
ds^2 = \frac{R^2}{z^2}[-h(z) dt^2 + f(z)^2 dz^2 + dx^2_1 + dx^2_2 + dx^2_3]~.
\end{eqnarray} 

\noindent This leads to the following area functional
\begin{eqnarray}
Area[\gamma_A] &=& R^3 \int_{-l/2}^{+l/2} dx_1 \frac{1}{z^3} \sqrt{1+(z^\prime f(z))^2} \int_{0}^{L}dx_2 \int_{0}^{L}dx_3 \nonumber\\
&=& 2 R^3 L^2  \int_{0}^{l/2} dx_1 \frac{1}{z^3} \sqrt{1+(z^\prime f(z))^2}~;~ z^\prime \equiv \frac{dz}{dx_1}~.
\end{eqnarray}
Once again we write down the area functional and the subsystem size $l$ in terms of the bulk coordinate $z$. This gives
\begin{eqnarray}\label{ym4}
Area[\gamma_A(z_*)] \equiv \mathcal{A}_l(z_*) &=& 2 R^3 L^2 \int_{a}^{z_*} \frac{z_*^3 f(z)}{z^3 \sqrt{z_*^6 - z^6}} dz\\
l &=& 2 \int_{0}^{z_*}  \frac{z^3 f(z)}{\sqrt{z_*^6 - z^6}} dz~.
\end{eqnarray}
\clearpage
\noindent The HEE is therefore given by
\begin{eqnarray}\label{ym1}
S_A = \frac{\mathcal{A}_l(z_*)}{4G_N} = \frac{R^3L^2}{2G_N} \int_{a}^{z_*} \frac{z_*^3 f(z)}{z^3 \sqrt{z_*^6 - z^6}} dz~.
\end{eqnarray}
From eq.(\ref{yangmills}), we can see that
\begin{eqnarray}\label{ym2}
\frac{dS_{EE}(l)}{dl} = 4\sqrt{\pi}\bigg[\frac{\Gamma(\frac{2}{3})}{\Gamma(\frac{1}{6})}\bigg]^3 \frac{N^2L^2}{l^3}
\end{eqnarray}
which gives using eq.(s)(\ref{equate}), (\ref{ym1}) and (\ref{ym2})
\begin{eqnarray}
l = 2 \sqrt{\pi} \bigg[\frac{\Gamma(\frac{2}{3})}{\Gamma(\frac{1}{6})}\bigg] z_*~.
\end{eqnarray}
Using the above relation, we can recast the expression of EE (\ref{yangmills}) in terms of the turning point $z_*$ as

\begin{eqnarray}\label{ym3}
S_{EE}(z_*) = \frac{R^3 L^2}{4 G_N a^2} - \frac{\sqrt{\pi} R^3 L^2}{4G_N}\bigg[\frac{\Gamma(\frac{2}{3})}{\Gamma(\frac{1}{6})}\bigg]\frac{1}{z_*^2}~~.
\end{eqnarray}

\noindent Now using eq.(s) (\ref{ym1}) and (\ref{ym3}), we obtain

\begin{eqnarray}\label{ym5}
\frac{R^3 L^2}{a^2} - \frac{\sqrt{\pi} R^3 L^2}{z_*^2}\bigg[\frac{\Gamma(\frac{2}{3})}{\Gamma(\frac{1}{6})}\bigg] = 2 R^3 L^2 \int_{a}^{z_*} \frac{z_*^3 f(z)}{z^3 \sqrt{z_*^6 - z^6}} dz~.
\end{eqnarray}

\noindent Solving the above integral equation yields
\begin{eqnarray}\label{ym6}
f(z)= \frac{3}{\pi} z^3 \frac{d}{dz} \bigints_{a}^{z} dz_* \frac{\bigg[ \frac{z_*^2}{a^2} - \sqrt{\pi}\bigg[\frac{\Gamma(\frac{2}{3})}{\Gamma(\frac{1}{6})}\bigg] \bigg]}{\sqrt{z^6-z_*^6}} \equiv \frac{3z^3}{\pi} \frac{dM}{dz}~.
\end{eqnarray}
We now proceed to compute $\frac{dM}{dz}$ in the above expression. This yields

\begin{eqnarray}\label{ym7}
\frac{dM}{dz} = \frac{\pi}{3z^3} - \sqrt{\pi}\bigg[\frac{\Gamma(\frac{2}{3})}{\Gamma(\frac{1}{6})}\bigg] \frac{(\frac{a}{z^4})}{\sqrt{1-(\frac{a}{z})^6}} - \frac{2 \sqrt{\pi} \Gamma(2/3)}{\Gamma(1/6)} (\frac{a}{z^4}) ~{}_2F_1\bigg[\frac{1}{6}, \frac{1}{2}, \frac{7}{6}, \bigg(\frac{a}{z}\bigg)^6\bigg]~. 
\end{eqnarray}
Substituting the above expression in eq.(\ref{ym6}), we obtain

\begin{eqnarray}\label{ym8}
f(z) = 1 - \frac{3}{\sqrt{\pi}}\bigg[\frac{\Gamma(\frac{2}{3})}{\Gamma(\frac{1}{6})}\bigg] \frac{(\frac{a}{z})}{\sqrt{1-(\frac{a}{z})^6}} - \frac{6}{\sqrt{\pi}}\bigg[\frac{\Gamma(\frac{2}{3})}{\Gamma(\frac{1}{6})}\bigg] \bigg(\frac{a}{z}\bigg) ~{}_2F_1\bigg[\frac{1}{6}, \frac{1}{2}, \frac{7}{6}, \bigg(\frac{a}{z}\bigg)^6\bigg]~.
\end{eqnarray}
It is remarkable to observe that the above expression for the metric coefficient $f(z)$ is exact upto all orders in $a/z$, $(z>a)$.
In the limit $\frac{a}{z} \rightarrow 0$, the above expression for $f(z)$ reduces to

\begin{eqnarray}
f(z) = 1~.
\end{eqnarray}
Now substituting $f(z)$ in eq.(\ref{ymmetric}), we get

\begin{eqnarray}
ds^2 = \frac{R^2}{z^2}\bigg[-h(z) dt^2 + dz^2 + dx^2_1 + dx^2_2 + dx^2_3\bigg] \equiv g_{\mu \nu} dx^{\mu} dx^{\nu}~;~\mu, \nu = 1,...,5~. 
\end{eqnarray} 
Using eq.(\ref{adsE}) in $4+1$-dimensions (which implies $\Lambda= -\frac{6}{R^2}$ in the Einstein's equation), the above metric leads to the following Einstein's field equations,

\begin{eqnarray}\label{ymadse}
G_{22} &=& - \frac{3 h^{\prime}(z)}{2zh(z)} = 0
\end{eqnarray}
\begin{eqnarray}\label{ymadse2}
G_{33} = G_{44} = G_{55} = -\frac{6 h(z) h^{\prime}(z) + z h^{\prime}(z)^2-2zh(z)h^{\prime \prime}(z)}{4zh(z)^2} = 0~.
\end{eqnarray}
Solving eq.(\ref{ymadse}), we obtain
\begin{eqnarray}
h(z)= Constant = \mathcal{D}
\end{eqnarray}
which satisfies eq.(\ref{ymadse2}). 
The exact form of the bulk geometry in the limit $\frac{a}{z} \rightarrow 0$ corresponding to $\mathcal{N} = 4$ super Yang-Mills theory therefore reads

\begin{eqnarray}
ds^2 = \frac{R^2}{z^2}\bigg[-\mathcal{D} dt^2 + dz^2 + dx^2_1 + dx^2_2 + dx^2_3\bigg]~.
\end{eqnarray}
This is the well known pure $AdS_5$ metric in Poincare coordinates.



\section{Conclusion}\label{conclusion}
In this paper, we briefly discuss the method of bulk reconstruction with the help of the holographic prescription of computing entanglement entropy (Ryu-Takayanagi prescription). The exact results of entanglement entropy in conformal field theory have been used to reconstruct the geometrical structure of the bulk metric. We consider planar symmetric, static asymptotically AdS metric in $(2+1)$-dimension in Poincare coordinates. The choice of the $2+1$-spacetime dimensions has been made since  the exact results of entanglement entropy of a conformal field theory are only available in $(1+1)$-dimensions. We start our analysis by obtaining the area functional corresponding to the static minimal surface ($\gamma_A$) and compute the entanglement entropy of the conformal field theory holographically using the Ryu-Takayanagi formula. 
We then compare this result with the exact result known from conformal field theoretic considerations.
The formalism is applied to the exact results of entanglement entropy corresponding to three different types of subsystems of the conformal field theory. The first scenario consists of obtaining the bulk metric corresponding to the conformal field theory on an infinite line. Remarkably we observe that the metric function can be calculated exactly upto all orders in $a/z$, $(z>a)$ thereby revealing the complete effect of the boundary UV cut-off `$a$' on the bulk metric. Once we have the exact form of the static metric, we obtain the metric coefficient corresponding to $dt^2$ by substituting the full metric ansatz (with the static sector now being reconstructed from the holographic prescription) in the Einstein's field equations with a cosmological constant $\Lambda = -\frac{1}{R^2}$. This then leads to the pure $AdS$ metric in Poincare coordinates. It is observed that away from the cut-off surface $z=a$ (that is in the direction towards the turning point), the dynamics of the boundary conformal field theory is holographically related to the gravitational theory with pure $AdS$ metric. We then carry out the same procedure for a conformal field theory on an infinite line at a finite temperature $T$. In this case the bulk reconstruction formalism leads to the BTZ black hole spacetime. It is quite remarkable that the BTZ black hole spacetime holographically emerges from a conformal field theory on an infinite line at a finite temperature.  Here also we have been successful in capturing the effect of the UV cut-off `$a$' on the bulk metric upto all orders in $a/z$. Further we note that the expression for the metric function $f(z)$ reduces to the corresponding result of the $1+1$-dimensional CFT on an infinite line at zero temperature, in the limit $\beta \rightarrow \infty$ (that is in the zero temperature limit), revealing the consistency of our analysis. We then carry our analysis for a conformal field theory on a circle. In this case also we are able to reconstruct the geometry using the holographic proposal and the circular symmetry of the problem. Finally, we repeat the investigation to obtain the $AdS_5$ geometry in the bulk, dual to the $\mathcal{N}=4$ super Yang-Mills theory living in $3+1$-dimensions. Our investigation in this paper once again confirms the validity of the AdS/CFT correspondence from the entanglement entropy view point. The current study can be extended to higher dimensions also if the exact results of the entanglement entropy from field theoretical considerations are known
as long as it avoids time-dependency and momentum in the describing the density matrix of the CFT. 

\section*{Acknowledgements}
A.S. would like to acknowledge the support by Council of Scientific and Industrial Research (CSIR, Govt. of India) for Junior Research Fellowship (CSIR File No.$09/106(0168)/2018$-EMR-$1$). S.G. would like to acknowledges the support by DST SERB under Start Up Research Grant (Young Scientist), File No.YSS/2014/000180. S.G. also acknowledges the support of the Visiting Associateship programme of IUCAA, Pune.

\section*{Author contribution statement}
All authors have contributed equally.

\end{document}